\documentclass[11pt]{article}
\usepackage{latexsym,amsmath,amsfonts,url,color,graphicx,subfigure,float}
\usepackage{float}
\newcommand{\ignore}[1]{}

\begin{document}
\begin{center}
{\LARGE\textbf{Opinion formation in multiplex networks with general initial distributions}}\\
\bigskip

Chris G. Antonopoulos$^{1*}$\qquad Yilun Shang$^{2\dagger}$

$^{1}$Department of Mathematical Sciences, University of Essex, Wivenhoe Park, UK\\

$^{2}$School of Mathematical Sciences, Tongji University, Shanghai, China\\
$^{*}$Email: \texttt{canton@essex.ac.uk}\\
$^{\dagger}$Email: \texttt{shyl@tongji.edu.cn}\\
\end{center}

\begin{abstract}

We study opinion dynamics over multiplex
networks where agents interact with bounded
confidence. Namely, two neighbouring individuals exchange
opinions and compromise if their opinions do not differ by more than
a given threshold. In literature, agents are generally assumed to
have a homogeneous confidence bound. Here, we study analytically and
numerically opinion evolution over structured networks characterised
by multiple layers with respective confidence thresholds and general
initial opinion distributions. Through rigorous probability
analysis, we show analytically the critical thresholds at
which a phase transition takes place in the long-term consensus behaviour, over multiplex networks with some regularity conditions. Our results
reveal the quantitative relation between the critical threshold and initial distribution. Further, our numerical simulations illustrate the consensus behaviour of the agents in network topologies including lattices and, small-world and scale-free networks, as well as for structure-dependent convergence parameters accommodating node heterogeneity. We find that the critical thresholds for consensus tend to agree with the predicted upper bounds in Theorems 4 and 5 in this paper. Finally, our results indicate that multiplexity hinders consensus formation when the initial opinion configuration is within a bounded range and, provide insight into information diffusion and social dynamics in multiplex systems modeled by networks.

\bigskip

\end{abstract}

The last decades witnessed many attempts to delineate the
propagation of opinions or behaviours in a structured population by
network science \cite{1}, where individuals are located on the node
set of a connected graph and characterised by their opinion. The
study of opinion dynamics covers a wide range of topics of interest,
such as collective decision-making, emergence of fads, minority
opinion survival, and emergence of extremism, etc., in the
communities of sociophysics, social simulation and complexity
science. Varied models have been developed to explain how
hierarchies \cite{2} and consensus \cite{3,4,4a} may arise in a
society. For more information and results in the broad field of
social dynamics, we refer the reader to the comprehensive surveys in
\cite{5,6}.

Due to the striking analogy with spin systems, the opinion models
with binary or discrete opinion space \cite{3,6} have dominated
research in the Physics' literature. In social contagion
processes, however, when people having opinions toward something
meet and discuss, they may adapt their opinions toward the other
individual's opinion and reach a compromise. In this context,
continuous opinion space with opinions expressed in real numbers
is more favourable since it allows adjustment in terms of averaging
due to the continuous nature of the opinions. Examples include
prices, tax rates or predictions about macroeconomic variables.
Following this paradigm, a well-known continuous-opinion model has
been proposed by Deffuant, Weisbuch, and others (Deffuant model)
\cite{7,8}, which further examines compromising agents under bounded
confidence. In such models, an individual is only willing to take
those opinions into account, which differ less than a certain bound
of confidence $d$ from their own. This assumption reflects the
psychological concept of selective exposure, where people tend to
avoid communication with those with conflicting opinions. Similar
consideration has been adopted in the much studied Axelrod model for
the dissemination of cultures \cite{9}.

In the initial studies of Deffuant-type opinion models, agents in a
network are assumed to be homogeneous and have the same confidence
bound. For instance, it was shown in \cite{7,10} that there exists a
universal critical confidence threshold $d_c$ for the homogeneous
Deffuant model, above which complete consensus is reached (namely, a
single opinion cluster emerges) while below it, opinions diverge
(namely, two or more opinion clusters are observed) through
extensive simulations on complex networks, be them complete graphs,
lattices, or scale-free networks. In recent years, agent-dependent
multi-level confidence bounds have been incorporated into the model,
which mirror the complicated physiological and psychological factors
such as the disparity of people's knowledge, experience, and
personality; see e.g. \cite{11,12,13,14}. The persuasion capacity of
the mass media has also been found to play a role in opinion
formation \cite{15}. It is worth noting that most of them are based on
numerical simulations with only a few exceptions \cite{16,17,18,29}
due to the complicated nonlinear dynamics involved.

The opinion negotiation processes studied in the above works take
place on networks containing edges of the same type and at the same
temporal and topological scale. However, the real individuals in a
society are usually simultaneously connected in multiple ways, which
can make a non-additive effect on network dynamics \cite{19,20}.
People in a society, for example, interact through diverse
relationships: friendship, partnership, kinship, vicinity,
work-related acquaintanceship, to name just a few. Admittedly, a
natural and more appropriate description of such systems can be
given by using multiplex networks, where the networks are made up of
different layers that contain the same nodes and a given type of
edges in each layer. Some recent works have pointed out that
multiplexity can result in intrinsically different dynamics from
their equivalent single-layer counterparts. The irreducibility of
the Ising and voter models on multiplex networks has been emphasized
in \cite{a1,a2}. Opinion competition dynamics on duplex networks has
been studied in \cite{a3}, where coexistence of both opinions in the
two layers has been found possible using mean-field approximation.
In the context of culture dissemination, multiplexity is found to
generate a qualitatively different dynamical behaviour for the Axelrod
model, which produces a new stable regime of cultural diversity
\cite{a4}. In addition to single information spreading process on
multiplex networks, the coupling between different types of contact
processes, such as opinion formation and disease spreading, has been
investigated in the multiplex networks; see e.g. \cite{22,23,24}.
Synchronisation processes between different layers in multiplex
networks featuring the interplay between distinctive topological
structures and dynamics have been reported recently in \cite{a5,a6}.
An updated survey towards the spreading processes and opinion
formation on multiplex networks can be found in \cite{21}. To the
best of our knowledge, little attention has been paid to the opinion
evolution in the Deffuant model (featuring bounded confidence) in the
context of multiplex networks. In \cite{25}, the author first
examined the Deffuant model in a multiplex network, which is modeled
by an infinite line with multiple layers. The critical confidence
threshold is analytically identified through probabilistic analysis
and verified by numerical simulations.

In this paper, we aim to moving a step further in the direction of
\cite{25} by considering both general initial opinion distributions
and general multiplex networks. In the standard Deffuant model, the
initial opinions are assumed to be independently and uniformly
distributed in the interval $[0,1]$. General initial distributions
have been independently introduced in \cite{18,26}. We first address
the opinion formation with general initial distributions over
1-dimensional multiplex networks after introducing our model in the
Model description section. We then generalise our results to
higher-dimensional multiplex lattices and, to general multiplex
networks satisfying some regularity conditions. We derive analytical
expressions for the critical confidence bound, where both the
structural multiplexity and the initial distribution play essential
role. Interestingly, we show that multiplexity essentially impedes
consensus formation in the situations when the initial opinion
configuration is within a bounded range. On the other hand, if a
substantial divergence exists in the initial opinions, whether it is
bounded distributed or not, multiplexity is found to play no role in
determining the critical confidence level. Extensive numerical
simulations are provided with both constant and degree-dependent
convergence parameters, and the paper is concluded with some open
problems in the Discussion section.

\section*{Methods}

\subsection*{Model description}

The class of models considered here are examples of
interacting particle systems \cite{27} combining features of
multiplex networks. Given $\ell\in\mathbb{N}$, a
multiplex network is a pair $G=(V,E)$, made of $\ell$ layers
$G_1,G_2,\ldots,G_{\ell}$ such that each layer is a simple graph
$G_i=(V,E_i)$ with node set $V$ and edge set $E_i\subseteq V\times
V$ for $i=1,\ldots, \ell$. Here, the node set $V$ is shared by all
layers and it can be either finite or infinite. The edge set of $G$
consists of $\ell$ types of edges: $E=\cup_{i=1}^{\ell}E_i$. From
the perspective of graph theory, each edge between two nodes $u$ and
$v$ in graph $G$ is a multiple edge consisting of at most $\ell$
parallel edges, each of which belongs to a respective layer $G_i$.
We assume that each layer $G_i$ has bounded degrees. Hence, each
agent in the network $G$ has a bounded number of neighbours and at
most $\ell$ types of relationship. Without loss of generality, we
may assume that the network $G$ is connected since one could
consider connected components separately in what follows.

In the Deffuant model \cite{7,8}, two agents compromise according to the
following rules: initially (at time $t=0$), each agent $u\in V$ is
assigned an opinion value $X_0(u)\in\mathbb{R}$ identically and
independently distributed (i.i.d.) following some distribution
$\mathcal{L}(X_0)$. In the standard case, $\mathcal{L}(X_0)$ is the
uniform distribution over $[0,1]$. Independently of this, in the $i$th
layer, each edge $e\in E_i$ is independently assigned a Poisson
process with rate $\lambda p_i$ with $p_i\in(0,1)$ and $\lambda>0$
for $i=1,\ldots,\ell$. We assume that $\sum_{i=1}^{\ell}p_i=1$
without loss of generality. These Poisson processes defined on the
edges in $E$ govern the evolution of opinions. Specifically, let
$X_t(u)$ be the opinion value of agent $u$ at time $t\ge0$, which
remains unchanged as long as no Poisson event happens for any
edge in $E$ incident to $u$. Let $d>0$, $\alpha_1=1$ and
$\alpha_i\in(0,1)$ for $i=2,\ldots,\ell$. When at some time $t$ the
Poisson event occurs at an edge $e=\{u,v\}\in E_i$ for some $i$,
such that the pre-meeting opinions of the two agents are
$X_{t-}(u):=\lim_{s\rightarrow t-}X_s(u)$ and
$X_{t-}(v):=\lim_{s\rightarrow t-}X_s(v)$, we set
\begin{equation}
X_t(u)=\left\{\begin{array}{cc}
X_{t-}(u)+\mu(X_{t-}(v)-X_{t-}(u)),&\mathrm{if}\
|X_{t-}(u)-X_{t-}(v)|\le \alpha_id;\\
X_{t-}(u),&\mathrm{otherwise},
\end{array}\right.\label{1}
\end{equation}
and
\begin{equation}
X_t(v)=\left\{\begin{array}{cc}
X_{t-}(v)+\mu(X_{t-}(u)-X_{t-}(v)),&\mathrm{if}\
|X_{t-}(u)-X_{t-}(v)|\le \alpha_id;\\
X_{t-}(v),&\mathrm{otherwise},
\end{array}\right.\label{2}
\end{equation}
where $\mu\in(0,1/2]$ is the so-called convergence parameter.
Therefore, if the two pre-meeting opinions lie at a distance less
than a certain confidence bound from one another, the meeting agents
will come closer to each other symmetrically, by a relative amount
$\mu$, where $\mu=1/2$ implies that the two agents meet halfway through. If not,
then they will stay unchanged. It is worth noting that the model is well-defined since the bounded degree assumption ensures that almost
surely (i.e., with probability 1) none of the Poisson events will be
simultaneous for an infinite node set \cite[p. 28]{27}.

The multiplexity in the above opinion model lies in two aspects.
First, the interaction rates $\lambda p_i$ in each layer can be
different. Second, the confidence bounds $\alpha_i d$ in each layer
can be different too. We might as well consider distinct convergence
parameters $\mu=\mu_i$ for the $i$th layer indicating different
willingness to change one's mind. However, it has been confirmed
analytically and numerically that $\mu$ plays no role in the
qualitative behaviour of the opinion dynamics; it rather only affects the
convergence time \cite{5,7,17,18}.

\subsection*{Sharing a drink process}

In the section, we briefly review the sharing a drink (SAD) process
proposed in \cite{17}, which is particularly useful in later
analysis of the Deffuant model on $\mathbb{Z}$; see also
\cite{13,18,26}.

Let $k\in\mathbb{N}\cup\{0\}$. The SAD process, denoted by
$\{Y_k(u)\}_{u\in\mathbb{Z}}$, is a deterministic process defined
iteratively as follows: set
\begin{equation}
Y_0(u)=\left\{\begin{array}{cl} 1,& \mathrm{for}\ u=0;\\
0,&\mathrm{for}\ u\in\mathbb{Z}\backslash\{0\}.
\end{array}\right.\label{A4}
\end{equation}
For a given sequence of nodes $u_1,u_2,\ldots\in\mathbb{Z}$ and
$\mu\in(0,1/2]$, we obtain the configuration
$\{Y_k(u)\}_{u\in\mathbb{Z}}$ for $k\ge1$ by setting
\begin{equation}
Y_k(u)=\left\{\begin{array}{cl}
Y_{k-1}(u)+\mu(Y_{k-1}(u+1)-Y_{k-1}(u)),& \mathrm{for}\ u=u_k;\\
Y_{k-1}(u)+\mu(Y_{k-1}(u-1)-Y_{k-1}(u)),& \mathrm{for}\ u=u_k+1;\\
Y_{k-1}(u),& \mathrm{for}\ u\in\mathbb{Z}\backslash\{u_k,u_k+1\}.
\end{array}\right.\label{SAD}
\end{equation}

This procedure can be envisioned as a liquid-exchanging process on
$\mathbb{Z}$. A glass is put at each site $u\in\mathbb{Z}$. At $k=0$
only the glass located at the origin is full (with value 1) while
all others are empty (with value 0). At each subsequent time step
$k$, one picks two neighbouring glasses at $u_k$ and $u_k+1$, and
pouring liquids from the glass with higher level to that with lower
level by a relative amount $\mu$. This gives rise to the SAD
process. The following lemma on unimodality can be easily proved.

\smallskip
\noindent\textbf{Lemma 1.} (Unimodality) \quad \itshape If
$u_j\not=-1$ for $j=1,\ldots,k$, then $Y_k(0)\ge Y_k(1)\ge
Y_k(2)\ge\ldots$. \normalfont
\smallskip

Fix $t>0$ and consider the opinion model on $\mathbb{Z}$. Note that
there exists a finite interval
$[u_\alpha,u_\beta]\subseteq\mathbb{Z}$ containing 0 such that the
Poisson events on the boundary edges $\{u_\alpha-1,u_\alpha\}\in
E_i$ and $\{u_\beta,u_\beta+1\}\in E_i$ for all $i=1,\ldots,\ell$
have not happened yet up to time $t$. Let $N$ be the number of
opinion adjustments that occur in $[u_\alpha,u_\beta]$ up to time $t$.
The times of these adjustments are arranged in the chronological
order
\begin{equation}
\tau_{N+1}:=0<\tau_N<\tau_{N-1}< \ldots<\tau_1\le t,\label{a1}
\end{equation}
where we set $\tau_{N+1}:=0$ for convenience. For $k=1,\ldots,N$, we
write $u_k$ as the left endpoint of the edge $\{u_k,u_k+1\}$ for
which $u_k$ and $u_k+1$ adjust opinions at time $\tau_k$. Given the
sequence $u_1,\ldots,u_N$ (in this order) and $\mu\in(0,1/2]$, we
obtain a SAD process $\{Y_k(u)\}_{u\in\mathbb{Z}}$ as defined by
(\ref{A4}) and (\ref{SAD}).

\smallskip
\noindent\textbf{Lemma 2.} (Linear representation) \quad \itshape
For $k=0,1,\ldots,N$,
\begin{equation}
X_t(0)=\sum_{u\in\mathbb{Z}}Y_k(u)X_{\tau_{k+1}}(u).\label{a2}
\end{equation}
Particularly,
$X_t(0)=\sum_{u\in\mathbb{Z}}Y_N(u)X_0(u):=\sum_{u\in\mathbb{Z}}Y_t(u)X_0(u)$.
\normalfont
\smallskip

This lemma implies that the constructed SAD process resembles the
dynamics of the corresponding Deffuant model backwards in time so
that the state $X_t(0)$ in the model at any time $t>0$ can be
expressed as a weighted average of states at time 0 with weights
given by an SAD configuration. See \cite{17,25} for a proof.

\section*{Results}

\begin{figure}[!ht]
    \centering\includegraphics[width=0.5\textwidth]{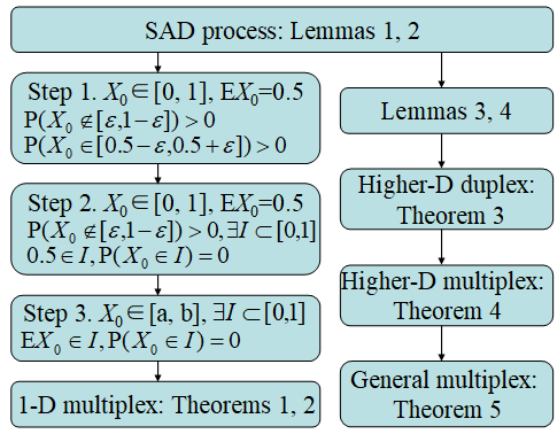}
\caption{\textbf{Schematic illustration of the theoretical results in
the paper.}}\label{fig_1}
\end{figure}

\subsection*{Opinion dynamics in 1-dimensional multiplex networks}

In this section, we will consider the multiplex opinion model on the
integers $\mathbb{Z}$, focusing on the general initial opinion
distributions. More specifically, we take $G=(V,E)$ with
$V=\mathbb{Z}$ and $E_i=\{\{u,u+1\}:u\in\mathbb{Z}\}$ for
$i=1,\ldots,\ell$. When $\ell=1$, $G$ becomes a simplex network with
only one type of edges. For this interaction network, the critical
confidence threshold for opinion formation with i.i.d. uniform
initial distribution in $[0,1]$ is $d_c=1/2$ \cite{16,17} and later extended to the multiplex 1-dimensional
networks in \cite{25}.

To appreciate this, we first present the results for the case
$\ell=2$ (see Theorem 1) and then extend it to the general multiplex
case (see Theorem 2), as illustrated in Fig. \ref{fig_1}. To this
end, we take $\ell=2$, $p=p_1$, and $\alpha=\alpha_2$. With these
assumptions, the main result concerning the critical confidence
threshold for the 1-dimensional duplex model reads as follows.

\smallskip
\noindent\textbf{Theorem 1.} (1-dimensional duplex networks)\quad
\itshape Consider the above continuous opinion model ($\ell=2$) on
$\mathbb{Z}$ with parameters $\lambda,d>0$, $\mu\in(0,1/2]$, and
$\alpha,p\in(0,1)$ with $\alpha>\mu$.

(a) Suppose that the initial opinion follows some bounded
distribution $\mathcal{L}(X_0)$ with expected value
$\operatorname{E}(X_0)$, whose support is contained in the smallest
closed interval $[a,b]$. Let $h\ge0$ be the length of some maximal
open interval $I\subset[a,b]$ satisfying $\operatorname{E}(X_0)\in
I$ and $\operatorname{P}(X_0\in I)=0$. Then,
$d_c=\max\{(\operatorname{E}X_0-a)(p+\alpha(1-p))^{-1},(b-\operatorname{E}X_0)(p+\alpha(1-p))^{-1},h\}$
is the critical confidence threshold in the following sense:
\begin{itemize}
\item If $d<\min\{d_c,b-a\}$, then with probability 1, there will be (infinitely
many) finally blocked edges, namely, $e=\{u,u+1\}$ satisfies
$|X_t(u)-X_t(u+1)|>d$ for all $t$ large enough;

\item If $d>\min\{d_c,b-a\}$, then with probability 1,
$X_{\infty}(u):=\lim_{t\rightarrow\infty}X_t(u)=\operatorname{E}(X_0)$
for every $u\in\mathbb{Z}$.
\end{itemize}

(b) Suppose that the initial opinion distribution $\mathcal{L}(X_0)$
is unbounded but its expectation exists in the sense of
$\operatorname{E}(X_0)\in\mathbb{R}\cup\{\pm\infty\}$. Then, for any
$d>0$, with probability 1, there will be (infinitely many) finally
blocked edges, namely, $e=\{u,u+1\}$ satisfying $|X_t(u)-X_t(u+1)|>d$
for all $t$ large enough. \normalfont
\smallskip

Before proceeding with the proof, we provide a couple of remarks.
Firstly, when the initial distribution $\mathcal{L}(X_0)$ is bounded
and $d<\min\{d_c,b-a\}$, we will show that
$\{|X_t(u)-X_t(u+1)|\}\in\{0\}\cup[d,b-a]$ for sufficiently large
$t$ and all $u\in\mathbb{Z}$, and hence, the integers split into
(infinitely many) finite clusters of neighbouring agents
asymptotically agreeing with each other, with no global consensus achieved. Secondly, in the special case of $\mathcal{L}(X_0)$ being
the standard uniform distribution in $[0,1]$, we readily reproduce
Theorem 1 in \cite{25}. A general $\mathcal{L}(X_0)$ has been
considered both theoretically and via simulations in \cite{18,26} for simplex networks (i.e., $\ell=1$). Theorem 1 can be thought of
as an extension to multiplex networks. Finally, the assumption
$\alpha>\mu$ is required here for technical reasons as in \cite{25},
which does not have a counterpart in the case of simplex networks where
$\mu$ only influences the convergence time of the negotiation
process.

The crucial technique adopted here is the SAD process introduced in \cite{17}. The SAD process and its basic
properties are briefly reviewed in the Method section. Another key
concept from that paper is the flat-points concept. To accommodate
the general distributions considered in the present paper, a slight
extension of the definitions therein can be provided as follows.
Given $\varepsilon>0$ and the initial opinion configuration
$\{X_0(v)\}_{v\in\mathbb{Z}}$ with finite expectation, a node
$u\in\mathbb{Z}$ is said to be an {\it $\varepsilon$-flat point to
the right} if for all $n\ge0$,
\begin{equation}
\frac{1}{n+1}\sum_{v=u}^{u+n}X_0(v)\in\left[\operatorname{E}(X_0)-\varepsilon,\operatorname{E}(X_0)+\varepsilon\right].\label{a3}
\end{equation}
Likewise, $u\in\mathbb{Z}$ is said to be an {\it $\varepsilon$-flat
point to the left} if for all $n\ge0$,
\begin{equation}
\frac{1}{n+1}\sum_{v=u-n}^{u}X_0(v)\in\left[\operatorname{E}(X_0)-\varepsilon,\operatorname{E}(X_0)+\varepsilon\right],\label{a4}
\end{equation}
and {\it two-sided $\varepsilon$-flat point} if for all $n,m\ge0$,
\begin{equation}
\frac{1}{n+m+1}\sum_{v=u-n}^{u+m}X_0(v)\in\left[\operatorname{E}(X_0)-\varepsilon,\operatorname{E}(X_0)+\varepsilon\right].\label{a5}
\end{equation}
We also define that $u\in\mathbb{Z}$ is an {\it $\varepsilon$-flat
point to the right at time $t$} if for all $n\ge0$,
$\frac{1}{n+1}\sum_{v=u}^{u+n}X_t(v)\in\left[\operatorname{E}(X_0)-\varepsilon,\operatorname{E}(X_0)+\varepsilon\right].$
Similar definitions for {\it $\varepsilon$-flat point to the left at
time $t$} and {\it two-sided $\varepsilon$-flat point at time $t$}
can be given.

\noindent\textbf{Proof of Theorem 1.} (a) Along the lines in
\cite[Section 2]{26}, we divide the proof of statement (a) into
three steps.

\textit{Step 1}. Suppose that the initial opinion distribution
$\mathcal{L}(X_0)$ is confined in [0,1] with expected value
$\operatorname{E}(X_0)=1/2$. Moreover, for any $\varepsilon>0$, we
assume that
$\operatorname{P}(X_0\not\in[\varepsilon,1-\varepsilon])>0$ and
$\operatorname{P}(1/2-\varepsilon\le X_0\le1/2+\varepsilon)>0$ hold.
Then we claim that $d_c=[2(p+\alpha(1-p))]^{-1}$ is the critical
confidence threshold in the same sense as in Theorem 1(a) (with $a=0$
and $b=1$).

To prove this claim, we need to show that the essential ingredients
in the proof of Theorem 1 in \cite{25} still hold true. We mention
here an obvious correction that the critical threshold separating
the subcritical and supercritical regimes therein should be
$\min\{d_c,1\}$ instead of $d_c$. For the subcritical regime, note
that the fact that the mass is around the expected value, i.e.,
$\operatorname{P}(1/2-\varepsilon\le X_0\le1/2+\varepsilon)>0$,
implies that $\operatorname{P}(u\ \mathrm{is}\
\varepsilon\mbox{-}\mathrm{flat}\ \mathrm{to}\ \mathrm{the}\
\mathrm{right})=\operatorname{P}(u\ \mathrm{is}\
\varepsilon\mbox{-}\mathrm{flat}\ \mathrm{to}\ \mathrm{the}\
\mathrm{left})$ $>0$ for all $\varepsilon>0$ and $u\in\mathbb{Z}$ by
similarly applying the coupling trick and the strong law of large
numbers. At time $t$ when a Poisson event occurs, define a Boolean
random variable $A_t$ by $A_t=1$ with probability $p$ and
$A_t=\alpha$ with probability $1-p$ so that the opinion model
constitutes a marked Poisson process with rate $\lambda$ \cite{25}.
We can then mimic the proof for Propositions 1 and 2 in \cite{25}
verbatim, which employs the condition
$\operatorname{P}(X_0\not\in[\varepsilon,1-\varepsilon])>0$ for any
$\varepsilon>0$.

For the supercritical regime, we need to note that the property
$\operatorname{P}(u\ \mathrm{is}\
\mathrm{two}\mbox{-}\mathrm{sided}\
\varepsilon\mbox{-}\mathrm{flat})>0$ for any $\varepsilon>0$ and
$u\in\mathbb{Z}$ can now be established by keeping in mind that
$\operatorname{P}(1/2-\varepsilon\le X_0\le1/2+\varepsilon)>0$
following the same reasoning as in \cite{17}; see also \cite{26}. Now
the proof for the supercritical regime in \cite{25} can be used,
which concludes the proof of \textit{Step 1}.

\textit{Step 2}. Suppose that the initial opinion distribution
$\mathcal{L}(X_0)$ is again confined in [0,1] with expected value
$\operatorname{E}(X_0)=1/2$.  For any $\varepsilon>0$, as in
\textit{Step 1} we assume that
$\operatorname{P}(X_0\not\in[\varepsilon,1-\varepsilon])>0$.
Moreover, assume that there exists some maximal open interval
$I\subset[0,1]$ of length $h$ satisfying $1/2\in I$ and
$\operatorname{P}(X_0\in I)=0$. Then, we claim that
$d_c=\max\{[2(p+\alpha(1-p))]^{-1},h\}$ is the critical confidence
threshold in the same sense of Theorem 1(a) (with $a=0$ and $b=1$).

When $d<h$, thanks to the assumption
$\operatorname{P}(X_0\not\in[\varepsilon,1-\varepsilon])>0$, we have
initial opinions both below and above $1/2$ with probability 1. Therefore, any edges which are blocked due to
initial incident opinions lying on different sides of the gap $I$
will remain blocked for all $t$. By ergodicity, there will be
infinitely many such blocked edges, and thus consensus can not be reached in
this case.

When $d>h$, we need to show that
\begin{equation}
\operatorname{P}(u\ \mathrm{is}\ \varepsilon\mbox{-}\mathrm{flat}\
\mathrm{to}\ \mathrm{the}\ \mathrm{right}\ \mathrm{at}\
\mathrm{time}\ t)=\operatorname{P}(u\ \mathrm{is}\
\varepsilon\mbox{-}\mathrm{flat}\ \mathrm{to}\ \mathrm{the}\
\mathrm{left}\ \mathrm{at}\ \mathrm{time}\ t)>0\label{3}
\end{equation}
for all $\varepsilon>0$, $u\in\mathbb{Z}$ and for some sufficiently
large $t$, since an arbitrary flat point at time $t=0$ no longer exists
due to the gap. Following the reasoning of \cite[Section 2]{26}, one can then
establish Eq. (\ref{3}). The only minor change that has to be made in
order to accommodate the multiplexity is that the involved marked
Poisson processes has rate $\lambda p+\lambda(1-p)=\lambda$ instead
of a unit rate, which does not affect the validity of the proof. Now, as
in \textit{Step 1}, we can mimic the proof of Propositions 1 and 2
in \cite{25} verbatim to settle the subcritical case. Accordingly,
$d_c\ge\max\{[2(p+\alpha(1-p))]^{-1},h\}$. Next, the two-sided
$\varepsilon$-flatness at time $t$ for any $\varepsilon>0$ can be
established similarly as in \cite[Section 2]{26}. Hence, the
argument in the supercritical case in \textit{Step 1} leads to
$d_c=\max\{[2(p+\alpha(1-p))]^{-1},h\}$, completing the proof of
\textit{Step 2}.

\textit{Step 3}. Now, everything is in place to prove Theorem 1(a) in
its full generality.

Define $c:=\max\{\operatorname{E}X_0-a,b-\operatorname{E}X_0\}$ and
perform the linear transformation $x\mapsto
(x-\operatorname{E}X_0)/2c+1/2$ to the dynamics of our multiplex
Deffuant model. Using the result in \textit{Step 2} and the fact
that the dynamics stays unchanged with respect to translations of
the initial distribution and that parameter $d$ can be re-scaled
as per a scaling transformation of the initial distribution in order
to recover the identical dynamics, we have
\begin{align}
d_c=&2c
\max\{[2(p+\alpha(1-p))]^{-1},h/2c\}\nonumber\\
=&\max\{(\operatorname{E}X_0-a)(p+\alpha(1-p))^{-1},(b-\operatorname{E}X_0)(p+\alpha(1-p))^{-1},h\}.\label{a6}
\end{align}
One can see that the ultimate consensus value in
the supercritical regime is transformed from $1/2$ to
$\operatorname{E}X_0$ in view of \textit{Step 2}.

(b) In the case of unbounded $\mathcal{L}(X_0)$, we divide the proof
into two cases.

\textit{Case 1.} $\operatorname{E}|X_0|<\infty$.

The strong law of large numbers implies that
\begin{equation}
\operatorname{P}\left(\frac{1}{n+1}\sum_{v=u}^{u+n}X_0(v)=\operatorname{E}X_0\right)=1\label{a7}
\end{equation}
for any $u\in\mathbb{Z}$. A simple calculation shows that node $u$
is $\delta$-flat to the right with positive probability for some
$\delta>0$.

Fix $d>0$. Following the reasoning in \cite[Proposition 1]{25} and
noting that $A_td\le d$, we can show that if $u-1$ and $u+1$ are
$\delta$-flat to the left and right respectively and
$X_0(u)\not\in[\operatorname{E}X_0-\delta-d,\operatorname{E}X_0+\delta+d]$
(which happens with positive probability), then $X_t(u-1)$ and
$X_t(u+1)$ will stay in the interval
$[\operatorname{E}X_0-\delta,\operatorname{E}X_0+\delta]$ for all
$t$ leaving the two edges $\{u-1,u\}$ and $\{u,u+1\}$ finally
blocked. Since this event happens at each $u\in\mathbb{Z}$ with
positive probability, it happens with probability 1 at infinitely
many nodes by ergodicity.

\textit{Case 2.} $\operatorname{E}X_0\in\{\pm\infty\}$.

Without loss of generality, we assume that
$\operatorname{E}X_0^+=\infty$ and $\operatorname{E}X_0^-<\infty$,
where $X_0^+$ and $X_0^-$ are the positive and negative parts of
$X_0$, respectively. We may further assume that
$\operatorname{P}(X_0\le0)>0$, otherwise a translation would
transform the problem to this case (c.f. \textit{Step 3} above).

Fix $d>0$. The same argument in \cite[Section 2]{26} can be used to show that the event $\mathcal{E}:=\{(1/n)\sum_{v=u+1}^{u+n}X_0(v)>d,\
\mathrm{for}\ \mathrm{all}\ n\in\mathbb{N}\}$ for any
$u\in\mathbb{Z}$ happens with positive probability. Similarly, along the lines of \cite[Proposition 1]{25}, we obtain that if
$\mathcal{E}$ happens and $X_0(u)\le0$ (which happens with positive
probability), then $X_0(u+1)>d$ for all $t$. Namely, there will
never be an opportunity for node $u+1$ to average with $u$. The same
thing holds for $u-1$ by symmetry. Since the initial opinions are
i.i.d., with positive probability we have $X_0(u)\le0$ and
$X_0(u-1),X_0(u+1)>d$, leaving the edges $\{u-1,u\}$ and $\{u,u+1\}$
finally blocked. Since this happens at every $u\in\mathbb{Z}$ with
positive probability,  by ergodicity, it happens with probability 1 at infinitely
many nodes. $\Box$

For a multiplex network $\mathbb{Z}$ with $\ell$ layers, Theorem 2
is within easy reach by essentially using the same arguments as above.

\smallskip
\noindent\textbf{Theorem 2.} (1-dimensional multiplex
networks)\quad \itshape Consider the above continuous opinion model
on $\mathbb{Z}$ with parameters $\lambda,d>0$, $\mu\in(0,1/2]$, and
$p_i\in(0,1)$ for $i=1,\ldots,\ell$, $\alpha_i\in(0,1)$ for
$i=2,\ldots,\ell$ and $\alpha_1=1$ with $\alpha_i>\mu$ for all
$i$.

(a) Suppose that the initial opinion follows some bounded
distribution $\mathcal{L}(X_0)$ with expected value
$\operatorname{E}(X_0)$, whose support is contained in the smallest
closed interval $[a,b]$. Let $h\ge0$ be the length of some maximal
open interval $I\subset[a,b]$ satisfying $\operatorname{E}(X_0)\in
I$ and $\operatorname{P}(X_0\in I)=0$. Then,
$d_c=\max\{(\operatorname{E}X_0-a)(\sum_{i=1}^{\ell}p_i\alpha_i)^{-1},(b-\operatorname{E}X_0)(\sum_{i=1}^{\ell}p_i\alpha_i)^{-1},h\}$
is the critical confidence threshold in the following sense:
\begin{itemize}
\item If $d<\min\{d_c,b-a\}$, then with probability 1, there will be (infinitely
many) finally blocked edges, namely, $e=\{u,u+1\}$ satisfies
$|X_t(u)-X_t(u+1)|>d$ for all $t$ large enough;

\item If $d>\min\{d_c,b-a\}$, then with probability 1,
$X_{\infty}(u):=\lim_{t\rightarrow\infty}X_t(u)=\operatorname{E}(X_0)$
for every $u\in\mathbb{Z}$.
\end{itemize}

(b) Suppose that the initial opinion distribution $\mathcal{L}(X_0)$
is unbounded but its expectation exists in the sense of
$\operatorname{E}(X_0)\in\mathbb{R}\cup\{\pm\infty\}$. Then, for any
$d>0$, with probability 1, there will be (infinitely many) finally
blocked edges, namely, $e=\{u,u+1\}$ satisfies $|X_t(u)-X_t(u+1)|>d$
for all $t$ large enough. \normalfont
\smallskip

Several observations can be drawn from Theorem 2. Firstly, when the
initial opinion distribution $\mathcal{L}(X_0)$ follows the standard
uniform distribution in $[0,1]$, we recover the previous result
\cite[Theorem 2]{25}. Secondly, when $\mathcal{L}(X_0)$ is bounded,
since $\sum_{i=1}^{\ell}p_i=1$, we always have
$d_c\ge\max\{(\operatorname{E}X_0-a),(b-\operatorname{E}X_0),h\}$,
where the equality holds if and only if $\ell=1$ or
$h\ge\max\{(\operatorname{E}X_0-a)(\sum_{i=1}^{\ell}p_i\alpha_i)^{-1},(b-\operatorname{E}X_0)(\sum_{i=1}^{\ell}p_i\alpha_i)^{-1}\}$.
This indicates it is more difficult to reach agreement over
multiplex networks than simplex networks in general. When there is a large
$h$, the critical confidence threshold $d_c$ is dominated by
$h$ and is independent from the multiplexity; on the other hand, for
relatively small $h$, the threshold is determined in turn by both
the multiplexity and the initial distribution. When the initial
distribution $\mathcal{L}(X_0)$ is unbounded, consensus cannot be
reached regardless of the multiplexity. Thirdly, if there exists
some $k$ satisfying $p_k\gg p_j$ for all $j\not=k$, then
$d_c\approx\max\{(\operatorname{E}X_0-a)\alpha_k^{-1},(b-\operatorname{E}X_0)\alpha_k^{-1},h\}$
in the case of bounded $\mathcal{L}(X_0)$. This suggests that the
critical confidence is governed by a frequently interacted layer in
the underlying network as one would expect.

\subsection*{Opinion dynamics in general multiplex networks}

In this section, we deal with more general multiplex networks and adopt a similar strategy by first looking into a duplex model on higher-dimensional lattices, generalising it to multiplex models and discussig on further extensions.

Particularly, we take $G=(V,E)$ with $V=\mathbb{Z}^D$ for $D\ge2$
and $E_i$ consists of all edges in the $D$-dimensional lattice for
$i=1,\ldots,\ell$. When $\ell=1$, $G$ becomes a simplex network with
only one type of edges; see \cite[Section 3]{26}. For $\ell=2$, we
denote $p=p_1$ and $\alpha=\alpha_2$ as in the above section. The
main result in this duplex case reads as follows.

\smallskip
\noindent\textbf{Theorem 3.} (higher-dimensional duplex
networks)\quad \itshape Consider the above continuous opinion model
($\ell=2$) on $\mathbb{Z}^D$ with $D\ge2$, $\lambda>0$,
$\mu\in(0,1/2]$, and $\alpha,p\in(0,1)$ with $\alpha>\mu$.

If the initial opinion is distributed on $[a,b]$ with expected value
$\operatorname{E}(X_0)$ and
$d>\frac{1}{2}(\operatorname{E}|2X_0-a-b|+b-a)(p+\alpha(1-p))^{-1}$,
then with probability 1,
$\lim_{t\rightarrow\infty}|X_t(u)-X_t(v)|=0$ for all edges
$\{u,v\}\in E$. \normalfont
\smallskip

Unlike the 1-dimensional case, here we are only able to establish
an upper bound for the critical confidence level $d_c$. In fact, as
commented in Remark 3.5 in \cite{26}, the case of $D\ge2$ is much
more complicated then the 1-dimensional counterpart and it is even not clear
if there exists a critical $d_c$ separating the supercritical and
subcritical regimes since the ultimate consensus does not need be
monotonic with respect to $d$. Furthermore, we note that the
consensus result in Theorem 3 is weaker that in Theorems 1 and 2
(for the supercritical regime) in the sense that only the difference
between the opinions of two neighbouring individuals is required to
converge towards zero. It is to verify that this
is equivalent to the convergence of each individual's opinion in a
finite network. For infinite networks considered in this paper,
however, the picture is quite different as one may imagine a
situation where the opinion shows wave-like patterns on broader and
broader spatial scales with non-vanishing amplitude as time increases.

To prove Theorem 3, we first define the energy of node $u$ at time $t$
as $\mathcal{E}_t(u)=f(X_t(u))$, where
$f:[a,b]\rightarrow[0,\infty)$ is some convex function. Given an
edge $e=\{u,v\}\in E$, let $T$ be the sequence of arrival times of
the Poisson events at $e$. The accumulated energy loss along $e$ is
defined as
\begin{equation}
\mathcal{E}_t^{\dagger}(e):=\sum_{s\in
T\cap[0,t]}(\mathcal{E}_{s-}(u)+\mathcal{E}_{s-}(v)-\mathcal{E}_{s}(u)-\mathcal{E}_{s}(v)),\label{a8}
\end{equation}
which is nonnegative due to Jensen's inequality \cite{26}. At time
$t$, the total energy of node $u$ is defined as
$\mathcal{E}_t(u)+\frac12\sum_{e:e\sim
u}\mathcal{E}_t^{\dagger}(e)$, where $e\sim u$ means $u$ is an
end-point of $e$. Following the same argument of \cite[Lemma
3.2]{26} and noting that the number of Poisson rings on a single
edge in any time period of length $\varepsilon$ is a Poisson random
variable with parameter $\lambda\varepsilon$, we have the following
lemma.

\smallskip
\noindent\textbf{Lemma 3.} For any $u\in\mathbb{Z}^D$ and time
$t\ge0$,
$\operatorname{E}\left(\mathcal{E}_t(u)+\frac12\sum_{e:e\sim
u}\mathcal{E}_t^{\dagger}(e)\right)=\operatorname{E}\mathcal{E}_0(0)$.
\normalfont
\smallskip

This means that the total energy at any node is conserved during the
opinion exchange process.

\smallskip
\noindent\textbf{Lemma 4.} For the above duplex opinion model on
$\mathbb{Z}^D$ with $D\ge2$, $\lambda>0$, $\mu\in(0,1/2]$, and
$\alpha,p\in(0,1)$. Suppose $\alpha>\mu$. If $d\in(0,b-a]$, then
with probability 1 for every two neighbours $u,v\in\mathbb{Z}^D$,
either $|X_t(u)-X_t(v)|>A_td$ for all sufficiently large $t$ (i.e.,
$\{u,v\}$ is finally blocked), or
$\lim_{t\rightarrow\infty}|X_t(u)-X_t(v)|=0$. \normalfont
\smallskip

\noindent\textbf{Proof.} As commented in \cite{25}, in the following,
we will use $A$ instead of $A_t$. Choose the energy function
$f(x)=x^2$ and fix an edge $e=\{u,v\}$. Let $\delta>0$. When there
is a Poisson event at $e$ at time $t$ and $u,v$ exchange opinions,
energy to the amount of $2\mu(1-\mu)(X_{t-}(u)-X_{t-}(v))^2$ is lost
along the edge; see \cite{25,26}. Hence, if
$|X_{t-}(u)-X_{t-}(v)|\in(\delta,Ad]$, energy
$\mathcal{E}_t^{\dagger}(e)$ will increase by the amount of at least
$2\mu(1-\mu)\delta^2$. Thanks to the memoryless property, given
$|X_{s}(u)-X_{s}(v)|\in(\delta,Ad]$ at some time $s$, the first
Poisson event after time $s$ on an edge incident to either $u$ or
$v$ occurs at $e$ with probability $(4d-1)^{-1}$.

In view of the conditional Borel-Cantelli lemma \cite[Corollary
3.2]{28}, this will happen infinitely often with probability 1. If
$|X_{t}(u)-X_{t}(v)|\in(\delta,Ad]$ at some sufficiently large $t$,
then $\lim_{t\rightarrow\infty}\mathcal{E}_t^{\dagger}(e)=\infty$.
However, this is impossible since Lemma 3 yields
$\operatorname{E}(\mathcal{E}_t^{\dagger}(e))\le
2\operatorname{E}(\mathcal{E}_0(0))\le2\max\{a^2,b^2\}$. Thereby,
with probability 1, for all large enough $t$,
$|X_t(u)-X_t(v)|\in[0,\delta]\cup(Ad,b-a]$.

For small enough $\delta>0$, $|X_t(u)-X_t(v)|$ cannot jump back and
forth between $[0,\delta]$ and $(Ad,b-a]$ infinitely often. This is
because a single Poisson event cannot increase $|X_t(u)-X_t(v)|$ by
more than $\mu d$, which for sufficiently small $\delta$, is always
less than the span of the gap $(\delta,Ad]$ that needs to be crossed
due to $\mu<\alpha$. Since there are only countably many edges, the
proof of Lemma 4 is completed. $\Box$

\noindent\textbf{Proof of Theorem 3.} Fix some $d\ge(a+b)/2$. If
$e=\{u,v\}$ be a finally blocked edge, then the opinion of node $u$
must finally be located in one of the intervals $[a,b-Ad)$ or
$(a+Ad,b]$. It follows from Lemma 4 that this event holds almost
surely for any $u$ if there are finally blocked edges. Suppose that
there is an edge $e$ such that
\begin{equation}
\operatorname{P}(e\ \mathrm{is}\ \mathrm{finally}\
\mathrm{blocked})>0.\label{4}
\end{equation}
Following a similar argument as in \cite[Lemma 3.4]{26}, we obtain with
probability 1 that
$\liminf_{t\rightarrow\infty}|X_t(u)-(a+b)/2|-a-Ad\ge(a+b)/2$ for
all $u\in\mathbb{Z}^D$.

We choose the energy function $f(x)=|x-(a+b)/2|$. By Lemma 3 and
Fatou's lemma, we obtain
\begin{align}
a+[p+\alpha(1-p)]d-\frac{a+b}{2}\le&\operatorname{E}\left(\liminf_{t\rightarrow\infty}\mathcal{E}_t(u)\right)=\operatorname{E}\left(\liminf_{t\rightarrow\infty}\left|X_t(u)-\frac{a+b}{2}\right|\right)\nonumber\\
\le&\liminf_{t\rightarrow\infty}\operatorname{E}\left|X_t(u)-\frac{a+b}{2}\right|\nonumber\\
\le&\liminf_{t\rightarrow\infty}\operatorname{E}\left(\mathcal{E}_t(u)+\frac12\sum_{e:e\sim
u}\mathcal{E}_t^{\dagger}(e)\right)\nonumber\\
=&\operatorname{E}\left(\mathcal{E}_0(u)\right)=\operatorname{E}\left|X_0-\frac{a+b}{2}\right|.\label{a9}
\end{align}
Recall that the condition of Theorem 3 implies that
$d>\frac{1}{2}(\operatorname{E}|2X_0-a-b|+b-a)(p+\alpha(1-p))^{-1}$,
which leads to a contradiction. Hence, the assumption (\ref{4}) must
not be true. The proof then follows from applying Lemma 4. $\Box$

Theorem 3 can be directly extended to the multiplex setting for a
general $\ell\ge2$.

\smallskip
\noindent\textbf{Theorem 4.} (higher-dimensional multiplex
networks)\quad \itshape Consider the above continuous opinion model
on $\mathbb{Z}^D$ with $D\ge2$, $\lambda>0$, $\mu\in(0,1/2]$, and
$p_i\in(0,1)$ for $i=1,\ldots,\ell$, $\alpha_i\in(0,1)$ for
$i=2,\ldots,\ell$ and $\alpha_1=1$. Suppose $\alpha_i>\mu$ for all
$i$.

If the initial opinion is distributed on $[a,b]$ with expected value
$\operatorname{E}(X_0)$ and
$d>\frac{1}{2}(\operatorname{E}|2X_0-a-b|+b-a)(\sum_{i=1}^{\ell}p_i\alpha_i)^{-1}$,
then with probability 1,
$\lim_{t\rightarrow\infty}|X_t(u)-X_t(v)|=0$ for all edges
$\{u,v\}\in E$. \normalfont
\smallskip

Some remarks follow: firstly, it is easy to check that the
lattice $\mathbb{Z}^D$ in Theorem 4 can be extended to any infinite,
locally finite, transitive and amenable connected graph
$G_i=(V,E_i)$ for each $i=1,\ldots,\ell$ by using Zygmund's ergodic
theorem; c.f. \cite[Remark 3.6]{26}. Recall that a graph is locally
finite if every node in it has a finite degree. A graph $G=(V,E)$ is
transitive if for any pair of nodes $u$ and $v$ in it, there is an
automorphism $\varphi:V\rightarrow V$ such that $\varphi(v)=u$. A
graph $G=(V,E)$ is amenable if there exists a sequence $S_n\subseteq
V$ of finite sets satisfying
$\lim_{n\rightarrow\infty}|\partial_{E}S_n|/|S_n|=0$, where
$\partial_{E}S_n$ is the edge boundary of $S_n$. The following
result can be established.

\smallskip
\noindent\textbf{Theorem 5.} (general multiplex networks)\quad
\itshape Consider the above continuous opinion model, where each
layer $G_i=(V,E_i)$ ($i=1,\ldots,\ell$) is an infinite, locally
finite, transitive and amenable connected graph. Let $\lambda>0$,
$\mu\in(0,1/2]$, and $p_i\in(0,1)$ for $i=1,\ldots,\ell$,
$\alpha_i\in(0,1)$ for $i=2,\ldots,\ell$ and $\alpha_1=1$ with
$\alpha_i>\mu$ for all $i$.

If the initial opinion is distributed on $[a,b]$ with expected value
$\operatorname{E}(X_0)$ and
$d>\frac{1}{2}(\operatorname{E}|2X_0-a-b|+b-a)(\sum_{i=1}^{\ell}p_i\alpha_i)^{-1}$,
then with probability 1,
$\lim_{t\rightarrow\infty}|X_t(u)-X_t(v)|=0$ for all edges
$\{u,v\}\in E$. \normalfont
\smallskip

Secondly, note that
$\frac{1}{2}(\operatorname{E}|2X_0-a-b|+b-a)<b-a$ unless (i)
$\operatorname{P}(X_0\in\{a,b\})=1$ and (ii) $X_0$ is not constant
with probability 1. This indicates that the condition
$d>\frac{1}{2}(\operatorname{E}|2X_0-a-b|+b-a)(\sum_{i=1}^{\ell}p_i\alpha_i)^{-1}$
in Theorem 4 stand a good chance to be nontrivial even for multiplex
networks in most meaningful situations. Thirdly, we have assumed
throughout this paper that the initial opinions following
$\mathcal{L}(X_0)$ are i.i.d. However, Theorems 4 and 5 still hold
if the initial opinions are stationary and ergodic with respect to
the graph automorphisms because no other specific features of i.i.d.
variables are used in the above proof. Finally, it seems that agents
forming a multiplex network are more difficult to reach consensus
for the same reason as remarked for 1-dimensional multiplex
networks in the above section. Furthermore, as we have mentioned in
the beginning of this section, it is generally even not clear if we
can still speak of critical confidence level $d_c$ in
$D$-dimensional ($D>1$) multiplex networks and more general
multiplex networks.

\subsection*{Numerical results}

In this section, we conduct agent-based simulations on different
finite multiplex networks, including regular ones such as
$D$-dimensional lattices which can be viewed as a truncation of
$\mathbb{Z}^D$ in Theorem 4, and irregular ones such as small-world
and scale-free networks, which obviously violate the regularity
conditions in Theorem 5 and are prominent examples of non power-law
and power-law networks, respectively. Interestingly, we see that for
all such networks, the critical thresholds of consensus tend to
agree with the predicted upper bounds in Theorems 4 and 5 in the
special cases of uniform $X_0$ and some choices of Poisson rates
associated with the multiple layers.

\begin{figure}[!ht]
    \centering\includegraphics[width=1\textwidth]{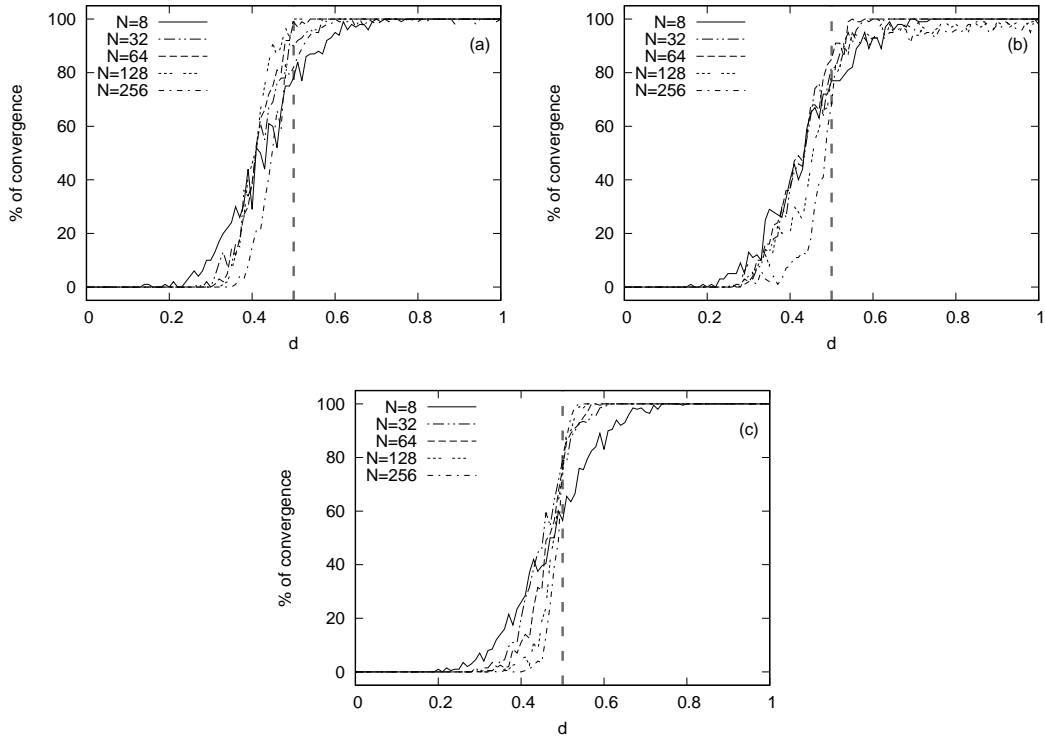}
\caption{\textbf{Percentage of opinion-convergence versus
$\boldsymbol{d}$ for constant $\boldsymbol{\mu}$ and, different
network sizes $\boldsymbol{N}$ and topologies.} Panel (a) is for
regular networks, panel (b) for Watts-Strogatz (small-world) and
panel (c) for Barab\'asi-Albert (scale-free) networks. The vertical
dashed line at $d=0.5$ corresponds to the point around which a
sudden jump occurs for increasing network sizes $N$. Note that in
these plots we have set the number of layers
$\ell$ to 4.}\label{fig_percent_convergence_vs_d}
\end{figure}

Particularly, in Fig. \ref{fig_percent_convergence_vs_d}, we plot
the percentage of convergence of opinions for five network sizes $N$
ranging from 8 to 256 with $\ell=4$ layers each, Poisson rate
$\lambda p_i=0.3$ and $\mu=0.5$ to maximise the convergence rate. At
$t=0$, we initialise each agent $u\in V$ by assigning an opinion
value $X_0(u)\in\mathbb{R}$ from the uniform distribution in
$(0,1)$. To check for convergence of opinions, we require that
$|X_t(u)-X_t(v)|<\hbar,\;\forall u,v\in G$, where $\hbar=10^{-5}$.
For each curve, we have run 100 simulations to compute the
percentage, each time for a different set of $\alpha$ values in Eqs.
\eqref{1} and \eqref{2}. Panel (a) shows the results for regular
lattices, whereas panel (b) for Watts-Strogatz (small-world) and (c) for Barab\'asi-Albert (scale-free) networks. In all cases,
we observe that the system reaches perfect consensus (i.e. 100\%
opinion convergence) or almost perfect consensus (i.e. $> 90\%$ opinion convergence), independently of the network structure, and
that this starts occurring for different $d$ values. It is worth
noting however that in all cases, convergence to consensus
is reached for $d>0.5$ denoted by the vertical dashed line in the plots.
Particularly, in panel (a), for $N=8$, the percentage of convergence
starts to increase from very small $d$ values with the tendency to
increase as $N$ increases, for example for $N=8$, it starts at
$d\approx 0.25$ whereas for $N=256$ at $d\approx0.37$. Surprisingly,
the jump from very small (almost 0\%) to very big (almost 100\%)
percentage of opinion convergence (reach of consensus) occurs at
$d=0.5$. Similar conclusions can be drawn for the Watts-Strogatz
(small-world) networks in panel (b) and Barab\'asi-Albert
(scale-free) networks in panel (c). This is reminiscent of a first
order phase-transition as a function of $d$ (order parameter) that might exist for infinitely big network sizes (i.e. $N\rightarrow\infty$) and is
an open question.

Note that the convergence parameter $\mu$ considered
above is constant and independent of the node-degrees and thus,
assumes that all agents have the same paces towards adjusting their
opinions. As this is rather ideal and not in agreement with real-life situations, we consider in the following two archetypal variants
given by
\begin{equation}
\mu^+(u,v)=\frac{\deg(u)\deg(v)}{2\max_{u\sim v}
\{\deg(u)\deg(v)\}}\label{a10}
\end{equation}
and
\begin{equation}
\mu^-(u,v)=\frac{\min_{u\sim v}
\{\deg(u)\deg(v)\}}{2\deg(u)\deg(v)},\label{a11}
\end{equation}
featuring the degree-centrality dependent scenarios,
where the max and min of $\deg(u)\deg(v)$ is computed among all
edges $u$, $v$ ($u\sim v$). Namely, we replace the convergence
parameter $\mu$ in Eqs. (\ref{1}) and (\ref{2}) by $\mu^+(u,v)$
indicating that higher-degree nodes are more willing to adjust their
opinions (and $\mu^-(u,v)$ indicating the opposite way). Clearly,
$\mu^+(u,v)$ and $\mu^-(u,v)$ are within the interval $(0,0.5]$ for
connected networks. $\mu^+$ is close to 0.5 for a well-connected
pair of nodes, whereas $\mu^-$ is close to 0.5 for a
poorly-connected pair of nodes. Our choice of degree-related
convergence parameter here naturally reflects the idea of the number of
neighbors/contacts in social networks, and models the possible
mechanisms of heterogeneous psychological, habitual and cultural
backgrounds in opinion spreading. Although degree-centrality
apparently depends on the structure of the layers of the multiplex
network, our numerical results indicate that the convergence
parameters do not affect the ultimate opinion configuration. It is
worth mentioning that there are other measures of centrality as well, such
as eigenvector-like centralities \cite{b4} and measures based on
random walks \cite{b5}, that have been studied in the context of multiplex
networks. However, as commented above, it is reasonable to expect similar
results for other convergence parameters mediated by more
complicated measures.

\begin{figure}[!ht]
    \centering\includegraphics[width=1\textwidth]{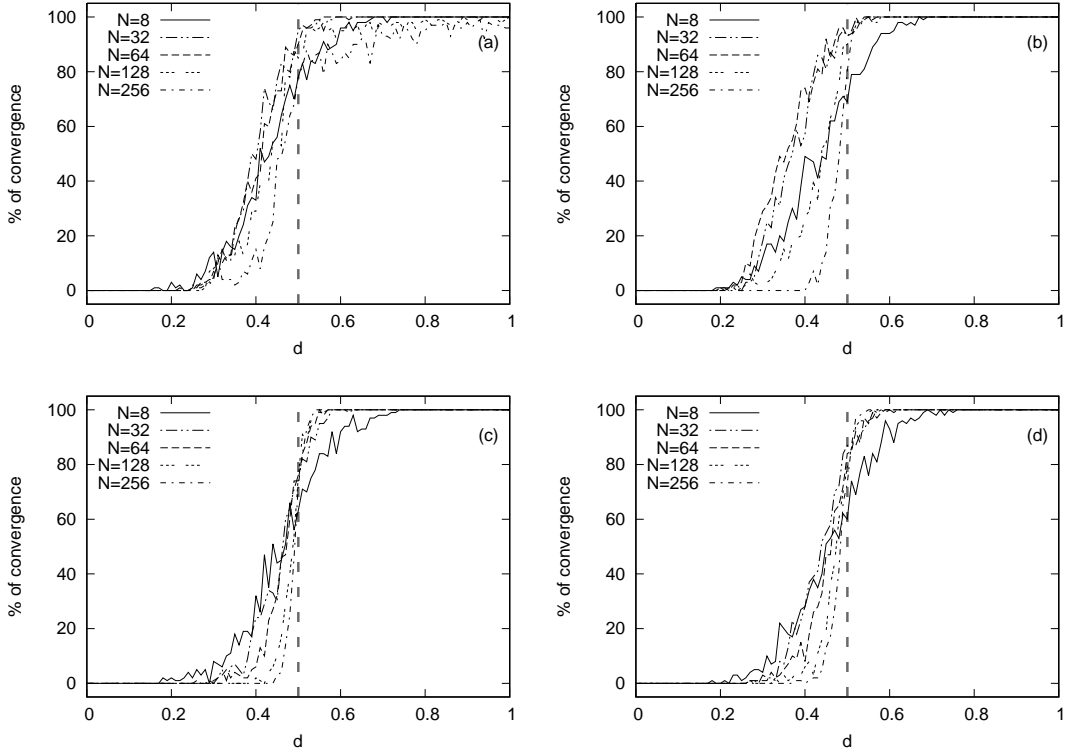}
\caption{\textbf{Percentage of opinion-convergence versus
$\boldsymbol{d}$ for different network sizes $\boldsymbol{N}$ and
topologies using the degree-dependent convergence parameters
$\boldsymbol{\mu^+(u,v)}$ and $\boldsymbol{\mu^-(u,v)}$ in Eqs.
(\ref{1}) and (\ref{2}).} Panels (a) and (b) are for Watts-Strogatz
(small-world) networks with $\mu^+(u,v)$ and $\mu^-(u,v)$,
respectively and panels (c) and (d) for Barab\'asi-Albert
(scale-free) networks with $\mu^+(u,v)$ and $\mu^-(u,v)$,
respectively. The vertical dashed line at $d=0.5$ corresponds to the
point around which a sudden jump occurs for increasing network sizes
$N$. Note that in these plots we have set the number of layers
$\ell$ to 4.}\label{fig_percent_convergence__mm_mp_vs_d}
\end{figure}

The simulation results for the degree-centrality dependent parameters $\mu^+$ and $\mu^-$ are presented in Fig.
\ref{fig_percent_convergence__mm_mp_vs_d}, where we have used the
same network sizes, number of layers, Poisson rate and number of
simulations as in Fig. \ref{fig_percent_convergence_vs_d} to compute
the percentage of convergence of the opinions. In particular,
panels (a) and (b) are for Watts-Strogatz (small-world)
networks with $\mu^+(u,v)$ and $\mu^-(u,v)$, respectively and panels
(c) and (d) for Barab\'asi-Albert (scale-free) networks with
$\mu^+(u,v)$ and $\mu^-(u,v)$, respectively. It is found that these
degree-dependent convergence parameters (i.e. $\mu^+(u,v)$ and
$\mu^-(u,v)$) do not alter the ultimate opinion configurations and that
the confidence threshold still remains. This is in line with previous
findings for Deffuant models in the case of single-layer networks
\cite{7,18} and our results extend this finding to multiplex networks.

\section*{Discussion}

In this paper, we studied analytically and numerically opinion dynamics over
multiplex networks with an arbitrary number of layers, where the
agents interact with each other with bounded confidence. In the
literature, agents are generally assumed to have a homogeneous
confidence bound and here we sought to study analytically and
numerically opinion evolution over multiplex networks with
respective confidence thresholds and general initial opinion
distributions. We explicitly identified the critical thresholds at
which a phase transition in the long-term consensus behaviour occurs.
We then discussed the interaction topology of the agents by
using multiplex $D$-dimensional lattices and extended them to general
multiplex networks under some regularity conditions. Our results
reveal the quantitative relation between the critical threshold and initial distribution. We also performed numerical
simulations and illustrated the consensus behaviour of the agents in
regular lattices and, small-world and scale-free networks. We found that the numerical results agree with our theoretical ones and
in particular, the critical thresholds of consensus tend to agree
with the predicted upper bounds in Theorems 4 and 5 for all
network topologies considered in the special cases of uniform $X_0$
and some choices of Poisson rates associated with the multiple
layers.

Moreover, we used the Deffuant opinion model represented as a
stochastic process for the evolution of opinions that includes
heterogenous confidence bounds and features general initial
distributions and, determined the critical threshold by employing
probability methods. The main results of our work are Theorems 2 (for $D=1$) and Theorem 4 (for $D>1$) which extend previous results in \cite{25,26} by considering
both multiplex structures with $\ell>1$ and general initial opinion
distribution $\mathcal{L}(X_0)$. We show that both the initial
distribution and multiplex structure play an important role in
the phase transition of opinion evolution in an infinite
$D$-dimensional regular lattice in the sense that the critical
confidence bound in the case of Theorem 2 (or an upper bound of it
in the case of Theorem 4) is influenced by both factors. Our results
indicate that multiplexity hinders consensus formation when the initial opinion configuration is within a
bounded range. This is numerically found to be true in more general
networks including small-world and scale-free networks,
which are ubiquitous in the real world. Our results provide insight
into information diffusion and social dynamics in multiplex
real-life systems modeled by networks. However, the theoretical proof of this
is beyond the scope of this paper as it would require the development of
substantially new techniques that we leave for a future publication.

It is worth mentioning that the networks considered
here are static, and thus the connectivity remains fixed throughout
opinion spreading. As a result, structural properties such as
centrality, correlations, homophily, and assortativity, remain the
same throughout opinion spreading. On the other hand, in networks of
human social interactions, the interaction can be assorted according
to, e.g., the channels used for communication such as face-to-face,
mobile phone, and social network services \cite{b1}. Certain social
mechanisms such as assortativity and homophily (namely, the tendency
of individuals to align to behaviours of their friends) are popular
in real social networks and may play a key role in opinion formation
and its dynamics. For instance, it is shown in \cite{b2} that the
higher the homophily between individuals in a multiplex network, the
quicker is the convergence towards cooperation in the social
dilemma. Multiplex social ecological network analysis unravels that
node heterogeneity has a critical effect on community robustness
\cite{b3}. However, as far as convergence of the opinion spreading
is concerned, our numerical results, for three different
characteristic types of multiplex networks (regular, small-world and
scale-free), indicate the same qualitative and almost similar
qualitative tendency to reach consensus as a function of $d$ for different network architectures. This is in agreement with our theoretical results. In
fact, assortativity and homophily are not included in our
theoretical analysis because they usually subvert the transitivity
and amenability conditions (see Theorem 5) that form the foundation
of our mathematical technique. In a future work, we will focus on how
to incorporate multiplex characterisations by means of structural
measures, such as homophily and assortativity of the multiplex
network, into analytically tractable opinion-formation models.
Finally, temporal or co-evolving networks with random environments
also seem appealing in this respect as they might lead to
differences with respect to convergence to consensus.

\section*{Acknowledgments}
Both authors would like to thank the support provided by the
University of Essex International Visiting Fellowships to the
project ``Opinion dynamics in multiplex networks''. Y. S. was partially
supported by the National Natural Science Foundation of
China (11505127) and the Shanghai Pujiang Program (15PJ1408300).

\section*{Competing financial interests}

The authors have no competing interest.

\section*{Data availability}

All data are generated by numerical simulations and they have all
been reported in the paper.

\section*{Author Contributions statement}

Y. S. devised the study. Y. S. and C. G. A. analysed the results and
wrote the main manuscript text. C. G. A. performed the numerical
analysis and generated the figures. Both authors reviewed the
manuscript.

\newpage

\section*{Figure legends}

\noindent\textbf{Fig. 1:} Schematic illustration of the theoretical results in
the paper.

\bigskip

\noindent\textbf{Fig. 2:} Percentage of opinion-convergence versus
$d$ for constant $\mu$ and, different
network sizes $N$ and topologies. Panel (a) is for
regular networks, panel (b) for Watts-Strogatz (small-world) and
panel (c) for Barab\'asi-Albert (scale-free) networks. The vertical
dashed line at $d=0.5$ corresponds to the point around which a
sudden jump occurs for increasing network sizes $N$. Note that in
these plots we have set the number of layers
$\ell$ to 4.

\bigskip

\noindent\textbf{Fig. 3:} Percentage of opinion-convergence versus
$d$ for different network sizes $N$ and
topologies using the degree-dependent convergence parameters
$\mu^+(u,v)$ and $\mu^-(u,v)$ in Eqs.
(\ref{1}) and (\ref{2}). Panels (a) and (b) are for Watts-Strogatz
(small-world) networks with $\mu^+(u,v)$ and $\mu^-(u,v)$,
respectively and panels (c) and (d) for Barab\'asi-Albert
(scale-free) networks with $\mu^+(u,v)$ and $\mu^-(u,v)$,
respectively. The vertical dashed line at $d=0.5$ corresponds to the
point around which a sudden jump occurs for increasing network sizes
$N$. Note that in these plots we have set the number of layers
$\ell$ to 4.

\end{document}